\documentclass[aps,prd,reprint,nofootinbib,superscriptaddress,floatfix]{revtex4-2}

\usepackage{amsmath,amssymb,bm}
\usepackage{graphicx}
\usepackage{booktabs}
\usepackage{microtype}
\usepackage{hyperref}
\hypersetup{colorlinks=true,linkcolor=blue,citecolor=blue,urlcolor=blue}

\begin{document}

\title{Generalized Regular Black Holes with Tunable Cores and Geodesic Properties}

\author{Hassan Hassanabadi}
\email{hassanhassanabadi@mail.fresnostate.edu}
\affiliation{Physics Department, California State University, Fresno, CA 93740, USA}

\begin{abstract}
We introduce a family of static and spherically symmetric regular black-hole geometries characterized by a discrete core index $n$, a regularization scale $a$, and a dimensionless deformation parameter $\eta$. The construction starts from a positive, normalized effective density profile with a tunable central behavior. Integrating the $tt$ component of Einstein's equations determines the mass function and hence the lapse, while the remaining Einstein equations fix the anisotropic pressures. The lowest member contains the Bardeen geometry as a limiting case, whereas a nonzero deformation parameter generates a Reissner--Nordstr\"om-like asymptotic correction. The parameter $n$ controls the order of central depletion: the lowest member possesses a de Sitter-type core, whereas higher members approach a Minkowski-like center with vanishing density and curvature. We determine the horizon structure, extremality condition, Hawking temperature, entropy, and the horizon thermodynamic identity implied by the radial Einstein equation. Null geodesics are analyzed to obtain the photon sphere, critical impact parameter, and shadow radius, including analytic weak-deformation approximations, the exact Schwarzschild and Bardeen limits, and comparison with the leading Reissner--Nordstr\"om-like behavior. The photon-orbit angular frequency and Lyapunov exponent are then used to separate orbital and instability timescales; their ratio provides information not already contained in the shadow radius. Weak gravitational lensing and magnification are discussed through a controlled sector-isolated expansion of the asymptotic metric. We further study timelike geodesics, circular motion, the marginally bound orbit, and the innermost stable circular orbit, deriving exact implicit relations and perturbative expressions. The model provides a unified framework for investigating how a tunable regular core and an RN-like exterior deformation affect the thermodynamic and geodesic properties of regular black holes.
\end{abstract}

\maketitle

\section{Introduction}

Regular black holes provide an important phenomenological arena for investigating whether the central singularity of the classical Schwarzschild and Kerr geometries can be replaced by a nonsingular high-curvature region while preserving the standard black-hole behavior at large distances. The subject originated with the Bardeen proposal and was subsequently developed through vacuum-like interiors, limiting-curvature constructions, and de Sitter-core geometries \cite{Bardeen1968,PoissonIsrael1988,FrolovMarkovMukhanov1990,Dymnikova1992}. A major development was the realization that regular geometries can arise in Einstein gravity coupled to nonlinear electrodynamics, leading to several electrically and magnetically charged nonsingular solutions and systematic extensions of the Bardeen construction \cite{AyonBeatoGarcia1998,AyonBeatoGarcia1999a,AyonBeatoGarcia1999b,Bronnikov2001}. Regularization mechanisms have also emerged from renormalization-group improvements, noncommutative geometry, and effective quantum-gravity considerations \cite{BonannoReuter2000,Nicolini2006,Hayward2006,Platania2019}. Considerable attention has consequently been devoted to generalized regular metrics, including charged, rotating, and multi-parameter families, as well as models constrained directly by the weak and dominant energy conditions \cite{LemosZanchin2011,BambiModesto2013,BalartVagenas2014a,BalartVagenas2014b,NevesSaa2014,FanWang2016,Rodrigues2018}. More systematic analyses have emphasized that regularity should be examined not only through finite curvature scalars but also through the causal and geodesic structure, the physical properties of the effective source, and the behavior of the energy conditions \cite{Frolov2016,Maeda2022,Lan2023}. An alternative direction replaces the conventional de Sitter interior by an asymptotically Minkowski core, thereby allowing the effective density and curvature to vanish toward the center and introducing qualitatively different internal structures \cite{SimpsonVisser2020,SimpsonVisser2022}. Recent investigations have further explored analytic matter models, regular-center replacements, and effective-field-theory or pure-gravity realizations of nonsingular geometries \cite{LiLu2023,Bronnikov2024,BuenoCanoHennigar2025}. These developments indicate that regular black holes are more naturally viewed as broad families of effective geometries whose physical properties depend sensitively on the assumed mass profile, matter distribution, and characteristic regularization scales.

An especially useful extension of this program is to construct regular black holes whose central structure can be continuously or discretely controlled by one or more shape parameters. In generalized mass-function approaches, additional parameters can regulate the characteristic curvature scale, the localization of the effective density, the strength of the central suppression, and the transition between the core and the asymptotic region \cite{BalartVagenas2014a,BalartVagenas2014b,FanWang2016,Frolov2016,Rodrigues2018,NevesSaa2014}. Such freedom is physically relevant because different regularization mechanisms need not produce the same interior: the familiar Bardeen, Dymnikova, and Hayward geometries possess finite-density de Sitter-like cores, whereas other constructions permit depleted or asymptotically Minkowski cores \cite{Bardeen1968,Dymnikova1992,Hayward2006,SimpsonVisser2020,SimpsonVisser2022,Maeda2022,LiLu2023,Bronnikov2024}. Parametric core models therefore provide a useful way of studying how changes concentrated in the strong-field interior propagate into the horizon structure, energy conditions, thermodynamics, photon motion, and massive-particle dynamics \cite{Frolov2016,FanWang2016,SimpsonVisser2020,Maeda2022,Lan2023}. They also make it possible to distinguish parameters controlling the degree of central regularization from those governing the asymptotic deformation of the metric. Motivated by this viewpoint, we introduce below a source-driven generalized family in which a discrete parameter controls the order of the central depletion, while an independent deformation parameter governs the exterior falloff of the effective source. The construction therefore provides a simple framework in which de Sitter-type and Minkowski-like regular cores can be analyzed within the same family and their consequences for black-hole thermodynamics and geodesic observables can be compared systematically.

The motivation for the present construction is to obtain a regular black-hole family in which the central geometry can be adjusted independently of the leading exterior deformation, while keeping the source simple enough to analyze analytically. Instead of prescribing the lapse function directly, we adopt a source-driven strategy and begin with a positive, normalized density profile whose central power is controlled by a discrete index $n$. This makes it possible to recover a finite-density de Sitter core for the lowest member and progressively depleted, Minkowski-like cores for higher members. At the same time, an independent parameter $\eta$ modifies the asymptotic falloff of the source and produces an RN-like correction without forcing the full spacetime to coincide with the Reissner--Nordstr\"om solution. This separation allows us to study, within one framework, how central regularization and exterior deformation influence the horizon structure, energy conditions, thermodynamics, photon propagation, weak lensing, and the dynamics of massive particles. The construction is therefore intended as a controlled phenomenological model for isolating which observable effects are primarily associated with the core and which arise from the asymptotic sector.

The paper is organized as follows. In Sec.~II we construct the geometry from the effective density, derive the corresponding mass function and lapse from Einstein's equations, discuss the central and asymptotic behavior, and summarize the energy conditions. Section~III is devoted to the horizon structure, including the extremal configuration and the parameter-dependent upper bound on the core index $n$. In Sec.~IV we study the Hawking temperature, entropy, and the horizon thermodynamic identity following from the radial Einstein equation. Section~V analyzes null geodesics, the photon sphere, the critical impact parameter, and the black-hole shadow. Section~VI examines the photon-orbit angular frequency and Lyapunov instability, with particular emphasis on separating the effects of $n$ and $\eta$. Weak gravitational lensing and magnification are considered in Sec.~VII. In Sec.~VIII we investigate timelike geodesics, circular motion, the marginally bound orbit, and the innermost stable circular orbit. Finally, Sec.~IX summarizes the main results and conclusions.

\section{Geometry and effective matter source}

The aim of this section is to construct the geometry from an effective source rather than to infer the source only after fixing the lapse. We therefore adopt a density-first strategy within Einstein gravity. The source is required to satisfy four basic criteria: regularity at the center, finite total mass, recovery of the Bardeen profile in a distinguished limit, and a controllable large-radius tail capable of generating an RN-like exterior correction. The $tt$ Einstein equation is then integrated to obtain the Misner--Sharp mass function $m(r)$, after which the lapse follows from the standard relation $f=1-2m/r$. Finally, the remaining Einstein equations determine the radial and tangential pressures. This construction is phenomenological rather than unique: no microscopic matter Lagrangian is assumed, but the geometry and its complete effective stress tensor are required to satisfy Einstein's equations exactly. Throughout we use geometrized units $G=c=\hbar=k_{\rm B}=1$.

We consider the static and spherically symmetric line element
\begin{equation}
 ds^2=-f(r)dt^2+\frac{dr^2}{f(r)}+r^2\left(d\theta^2+\sin^2\theta\,d\phi^2\right),
\end{equation}
written in the one-function Schwarzschild gauge. In spherical symmetry it is convenient to introduce the Misner--Sharp mass function \cite{MisnerSharp1964} through
\begin{equation}
 f(r)=1-\frac{2m(r)}{r}.
\end{equation}
For an anisotropic effective source $T^{\mu}{}_{\nu}=\operatorname{diag}(-\rho,p_r,p_t,p_t)$, the quantity $\rho(r)=-T^{t}{}_{t}$ is the local energy density of the effective source supporting the geometry. In any static region, a static observer with four-velocity $u^{\mu}=f^{-1/2}(1,0,0,0)$ measures $\rho=T_{\mu\nu}u^{\mu}u^{\nu}$. We do not identify this source a priori with ordinary matter or with a specific fundamental field; rather, it is a phenomenological anisotropic stress tensor representing the effective physics responsible for regularizing the central region. Its large-radius behavior will provide an additional physical interpretation below. The $tt$ Einstein equation associated with Eqs.~(1)--(2) is then $m'(r)=4\pi r^2\rho(r)$. We now select the following family of normalized densities
\begin{equation}
 \begin{aligned}
 \rho(r)&=\frac{M(2n+1)a\left(a+\eta s\right)r^{2n-2}}
 {4\pi s\left(s+\eta a\right)^{2n+2}},\\
 s(r)&=\sqrt{r^2+a^2}.
 \end{aligned}
\end{equation}
Here $M>0$ is the total mass scale, $a>0$ is the regularization length, $\eta\ge0$ is dimensionless, and $n=1,2,3,\ldots$ is the discrete core index. The choice is guided by the stated physical criteria. The factor $r^{2n-2}$ gives a finite central density for $n=1$ and an increasingly depleted core for $n>1$, while $a$ fixes the regularization length. The factor $a+\eta s$ changes the large-radius decay from $r^{-5}$ at $\eta=0$ to $r^{-4}$ for $\eta>0$. This provides the RN-like asymptotic sector. All factors are non-negative for the parameter range above, so $\rho(r)\ge0$. The normalization is fixed so that the integrated density gives the finite ADM mass $M$. The profile is not claimed to be unique; it is a minimal analytic interpolation that realizes these requirements simultaneously.

\begin{figure}[t]
 \centering
 \includegraphics[width=0.96\columnwidth]{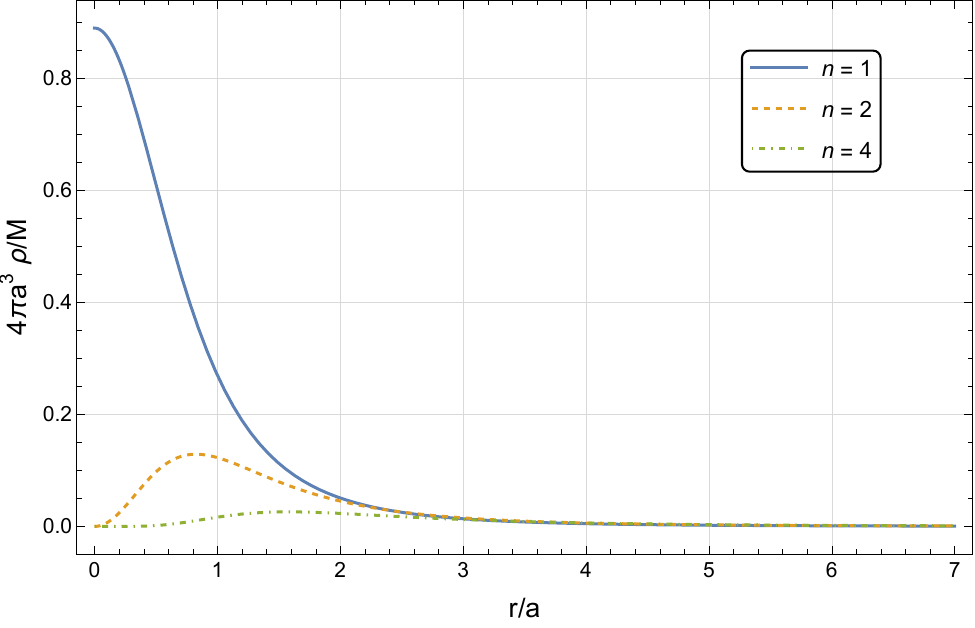}
 \caption{Dimensionless effective density $4\pi a^3\rho/M$ as a function of $r/a$ for $\eta=0.5$ and representative core indices. The $n=1$ profile remains finite at the origin, whereas the $n>1$ profiles vanish at the center and peak at finite radius, displaying the tunable transition from a de Sitter-type to a depleted core.}
 \label{fig:density}
\end{figure}

The role of the discrete index is illustrated in Fig.~\ref{fig:density}. We plot the dimensionless density $4\pi a^3\rho/M$ for a representative deformation $\eta=0.5$. The $n=1$ member has a finite central value, while the higher-$n$ profiles vanish at the origin and develop their maxima at finite radius. Thus increasing $n$ moves the effective support away from the center and makes the core increasingly depleted.

Indeed, the density is arranged so that $4\pi r^2\rho=d\{M[r/(s+\eta a)]^{2n+1}\}/dr$. Integrating the $tt$ equation from the regular center, with $m(0)=0$, therefore gives the mass function and the corresponding lapse in one step,
\begin{equation}
 \begin{aligned}
 m(r)&=M\left[\frac{r}{s+\eta a}\right]^{2n+1},\\
 f(r)&=1-\frac{2Mr^{2n}}{\left(s+\eta a\right)^{2n+1}}.
 \end{aligned}
\end{equation}
The normalization of the density is therefore not an independent assumption after the integration; it is equivalent to
\begin{equation}
 4\pi\int_{0}^{\infty}\rho(r)r^2dr=M,
 \qquad
 \lim_{r\to\infty}m(r)=M=M_{\rm ADM}.
\end{equation}
Thus the parameter $M$ appearing in the density is precisely the total gravitational mass measured at infinity.

The remaining Einstein equations determine the pressures. For the one-function metric they can be summarized as \cite{Dymnikova1992,FanWang2016}
\begin{equation}
 \begin{aligned}
 8\pi\rho&=\frac{2m'}{r^2},
 &8\pi p_r&=-\frac{2m'}{r^2},\\
 8\pi p_t&=-\frac{m''}{r},
 &p_r&=-\rho,
 \qquad p_t=-\rho-\frac{r}{2}\rho'.
 \end{aligned}
\end{equation}
Hence $p_r=-\rho$ is not imposed as a separate equation of state; it follows from the Schwarzschild gauge $g_{tt}g_{rr}=-1$. The conservation equation $\nabla_\mu T^{\mu r}=0$ is then satisfied identically by these relations.

It is useful at this point to summarize the energy conditions without introducing lengthy pressure expressions. Since $\rho\ge0$, define the logarithmic density slope
\begin{equation}
 \begin{aligned}
 q(r)\equiv-\frac{r\rho'}{\rho}
 =&-(2n-2)-\frac{\eta r^2}{s(a+\eta s)}+\frac{r^2}{s^2}\\
 &+\frac{(2n+2)r^2}{s(s+\eta a)}.
 \end{aligned}
\end{equation}
Equation (6) gives $p_t/\rho=-1+q/2$. The radial null condition is always saturated, $\rho+p_r=0$, while the tangential combination is $\rho+p_t=\rho q/2$. Consequently, the NEC and WEC reduce to $q\ge0$; the SEC requires $q\ge2$; and the DEC requires $0\le q\le4$, in the same spirit as energy-condition analyses of regular black-hole sources \cite{BalartVagenas2014a,Rodrigues2018}. These criteria make the parameter dependence transparent. For $n=1$, $q\simeq5(r/a)^2/(1+\eta)$ near the center, so the NEC, WEC, and DEC are satisfied there, whereas the SEC is violated as expected for a de Sitter-like core. For $n>1$, the density itself vanishes at the center and $q$ is understood through the limit $r\to0^+$, for which $q\to2-2n<0$. Hence the NEC/WEC and DEC are violated in a sufficiently small neighborhood of the depleted core; in a black-hole configuration this region may lie behind the event horizon. At large radius, $q\to4$ for every fixed $\eta>0$, so the NEC, WEC, and SEC hold asymptotically and the DEC is saturated. By contrast, on the $\eta=0$ Bardeen-type branch one has $q\to5$, so the tangential DEC is violated asymptotically. For any particular black-hole solution the physically relevant exterior test is therefore to evaluate $q(r)$ for $r\ge r_+$.

The small-$r$ behavior clarifies the geometric role of $n$. Expanding Eq.~(3) gives the central density
\begin{equation}
 \rho(0)=
 \begin{cases}
 \dfrac{3M}{4\pi a^3(1+\eta)^3}, & n=1,\\[1mm]
 0, & n>1.
 \end{cases}
\end{equation}
The corresponding central lapse follows from Eq.~(4),
\begin{equation}
 f(r)\simeq1-\frac{2M}{a^{2n+1}(1+\eta)^{2n+1}}r^{2n}.
\end{equation}
For $n=1$ this takes the standard de Sitter form
\begin{equation}
 \begin{aligned}
 f(r)&=1-\frac{\Lambda_{\rm eff}}{3}r^2+\mathcal O(r^4),\\
 \Lambda_{\rm eff}&=\frac{6M}{a^3(1+\eta)^3}=8\pi\rho(0).
 \end{aligned}
\end{equation}
Thus the first member has a finite-density de Sitter core. For every $n>1$ the $r^2$ term is absent and $f(r)=1-\mathcal O(r^{2n})$, producing an increasingly Minkowski-like core. The regularity can be checked directly from curvature invariants. Writing $C_n=2M/[a^{2n+1}(1+\eta)^{2n+1}]$, the central expansion $f=1-C_n r^{2n}+\cdots$ gives
\[
 \begin{aligned}
 R&=2(n+1)(2n+1)C_n r^{2n-2}+\cdots,\\
 K&=4(4n^4-4n^3+5n^2+1)C_n^2 r^{4n-4}+\cdots .
 \end{aligned}
\]
Here $K=R_{\mu\nu\rho\sigma}R^{\mu\nu\rho\sigma}$. For $n=1$, one obtains $R(0)=4\Lambda_{\rm eff}$ and $K(0)=8\Lambda_{\rm eff}^2/3$. For every $n>1$, both invariants vanish as $r\to0$. The analytic even-power expansion of $f(r)$ then ensures that the center is free of curvature singularities throughout the family.

The ordinary Bardeen geometry is recovered directly from the source-driven solution. Setting $n=1$ and $\eta=0$ in Eq.~(4) gives the original Bardeen form \cite{Bardeen1968}; its nonlinear-electrodynamics reinterpretation was later established in Ref.~\cite{AyonBeatoGarcia2000},
\begin{equation}
 m(r)=\frac{Mr^3}{(r^2+a^2)^{3/2}},
 \qquad
 f(r)=1-\frac{2Mr^2}{(r^2+a^2)^{3/2}}.
\end{equation}
The asymptotic region provides a second check. Expanding the mass function in Eq.~(4) for $r\gg a$ gives
\begin{equation}
\begin{aligned}
 m(r)=M\Bigg[&1-(2n+1)\eta\frac{a}{r}\\
 &+\frac{2n+1}{2}\left((2n+2)\eta^2-1\right)\frac{a^2}{r^2}
 +\mathcal O(r^{-3})\Bigg].
\end{aligned}
\end{equation}
Using $f=1-2m/r$ then yields
\begin{equation}
\begin{aligned}
 f(r)=&1-\frac{2M}{r}+\frac{2M(2n+1)\eta a}{r^2}\\
 &+\frac{M(2n+1)\left[1-(2n+2)\eta^2\right]a^2}{r^3}
 +\mathcal O(r^{-4}).
\end{aligned}
\end{equation}
The coefficient of the $1/r^2$ term defines the effective asymptotic charge scale
\begin{equation}
 Q_{\rm eff}^2=2M(2n+1)\eta a.
\end{equation}
Thus, for $\eta>0$, Eq.~(13) takes the RN-like form $f(r)=1-2M/r+Q_{\rm eff}^2/r^2+\mathcal O(r^{-3})$. The same interpretation follows from the source itself, because Eq.~(3) gives
\begin{equation}
 \rho\simeq\frac{Q_{\rm eff}^2}{8\pi r^4},
 \qquad p_r\simeq-\rho,
 \qquad p_t\simeq\rho.
\end{equation}
This is the algebraic stress-tensor structure of a radial Maxwell field. We emphasize that $Q_{\rm eff}$ is only an asymptotic geometric parameter; no microscopic electromagnetic field has been assumed. Finally, $a\to0$ in Eq.~(4) gives $f(r)\to1-2M/r$, so the Schwarzschild solution is recovered.

The $n=1$ member deserves a separate comment because it makes the role of $\eta$ especially transparent. At $\eta=0$ it is precisely the ordinary Bardeen geometry, whereas $\eta>0$ preserves the regular de Sitter-type center while lowering its central density and curvature scale through the factor $(1+\eta)^{-3}$. More importantly, switching on $\eta$ changes the asymptotic density from the Bardeen-type $r^{-5}$ falloff to an $r^{-4}$ tail and generates the RN-like scale $Q_{\rm eff}^{2}=6M\eta a$. Consequently, the asymptotic tangential DEC violation of the pure Bardeen branch, for which $q\to5$, is replaced by the limiting value $q\to4$, so the DEC is saturated at large radius. This improvement is asymptotic and should not be interpreted as proof that the DEC holds at every radius. As shown in the geodesic sections below, the same deformation also introduces corrections linear in $\eta a/M$ to the photon and circular-orbit scales, whereas the pure Bardeen corrections begin at quadratic order in $a/M$. Thus, for $n=1$, $\eta$ acts as a genuine exterior deformation of a regular Bardeen-like core rather than as a redundant parameter.

\section{Horizon structure}

Having established regularity, the ADM mass, and the asymptotic limits, we now turn to the causal structure of the geometry. In the coordinates of Eq.~(1), a Killing horizon occurs at a zero of $f(r)$, since both $g_{tt}$ and $g^{rr}$ vanish there. The number of positive roots distinguishes nonextremal black holes, extremal configurations, and horizonless geometries. Substituting the lapse (4) into $f(r_h)=0$ gives
\begin{equation}
 2Mr_h^{2n}=\left(s_h+\eta a\right)^{2n+1},
 \qquad s_h=\sqrt{r_h^2+a^2}.
\end{equation}
The same relation can be solved for the mass parameter,
\begin{equation}
 M(r_h)=\frac{(s_h+\eta a)^{2n+1}}{2r_h^{2n}},
\end{equation}
which is particularly useful when the horizon radius is used as the thermodynamic variable.

For a nonextremal black hole, Eq.~(16) may possess two positive roots. These roots coalesce when the minimum of the lapse just touches zero. The extremal configuration is therefore determined by the simultaneous conditions
\begin{equation}
 f(r_e)=0,
 \qquad
 f'(r_e)=0.
\end{equation}
To solve them analytically, we first use the horizon equation (16) to eliminate $M$ from $f'(r_e)=0$ and introduce the dimensionless quantity $y_e=s_e/a=\sqrt{1+r_e^2/a^2}$. The remaining condition reduces to the quadratic equation
\begin{equation}
 y_e^2-2n\eta y_e-(2n+1)=0.
\end{equation}
Equation (19) has the two algebraic roots
\begin{equation}
 y_e^{(\pm)}=n\eta\pm\sqrt{n^2\eta^2+2n+1}.
\end{equation}
For $n\ge1$ and $\eta\ge0$, the square root is strictly larger than $n\eta$, so $y_e^{(-)}<0$. On the other hand, by definition $y_e=\sqrt{1+r_e^2/a^2}\ge1$. The negative branch is therefore unphysical, leaving
\begin{equation}
 y_e=n\eta+\sqrt{n^2\eta^2+2n+1}.
\end{equation}
Once $y_e$ is known, the extremal radius follows directly from its definition,
\begin{equation}
 r_e=a\sqrt{y_e^2-1}.
\end{equation}
Finally, inserting Eq.~(22) into the horizon mass relation (17), or equivalently evaluating Eq.~(17) at $r_h=r_e$, gives the critical mass-to-length ratio
\begin{equation}
 \frac{M_e}{a}=\frac{(y_e+\eta)^{2n+1}}{2(y_e^2-1)^n}.
\end{equation}
Equations (21)--(23) therefore follow from the two extremality equations (18): Eq.~(21) selects the physical root, Eq.~(22) translates the dimensionless root back to a radius, and Eq.~(23) uses the horizon equation to determine the associated mass.

The horizon classification is consequently global rather than merely local. The function $M(r_h)$ in Eq.~(17) diverges as $r_h\to0^+$ and as $r_h\to\infty$, while the extremality equation has only one physical stationary point. Hence $M>M_e$ gives two horizons, $M=M_e$ one degenerate horizon, and $M<M_e$ no horizon. Regularity itself does not impose a finite upper bound on $n$, but the requirement that the spacetime actually represent a black hole does. For fixed $M/a$ and $\eta$, the admissible integers must satisfy
\begin{equation}
 \frac{M}{a}\ge\frac{M_e(n,\eta)}{a},
\end{equation}
and hence one may define $n_{\max}$ as the largest positive integer satisfying Eq.~(24). This makes the upper bound on $n$ a horizon-existence constraint rather than an assumption about the matter profile. The large-$n$ behavior makes the restriction explicit. On the Bardeen-type branch,
\[
 \left(\frac{M_e}{a}\right)_{\eta=0}=\frac{(2n+1)^{(2n+1)/2}}{2(2n)^n}\sim\sqrt{\frac{en}{2}},
\]
whereas for any fixed $\eta>0$, $M_e/a\sim e\eta n$. Thus, at large $n$, one has the estimates $n_{\max}\sim2(M/a)^2/e$ for $\eta=0$ and $n_{\max}\sim(M/a)/(e\eta)$ for fixed nonzero $\eta$. These are asymptotic estimates; the exact admissible integer is always determined from Eq.~(24).

\begin{figure}[t]
 \centering
 \includegraphics[width=0.96\columnwidth]{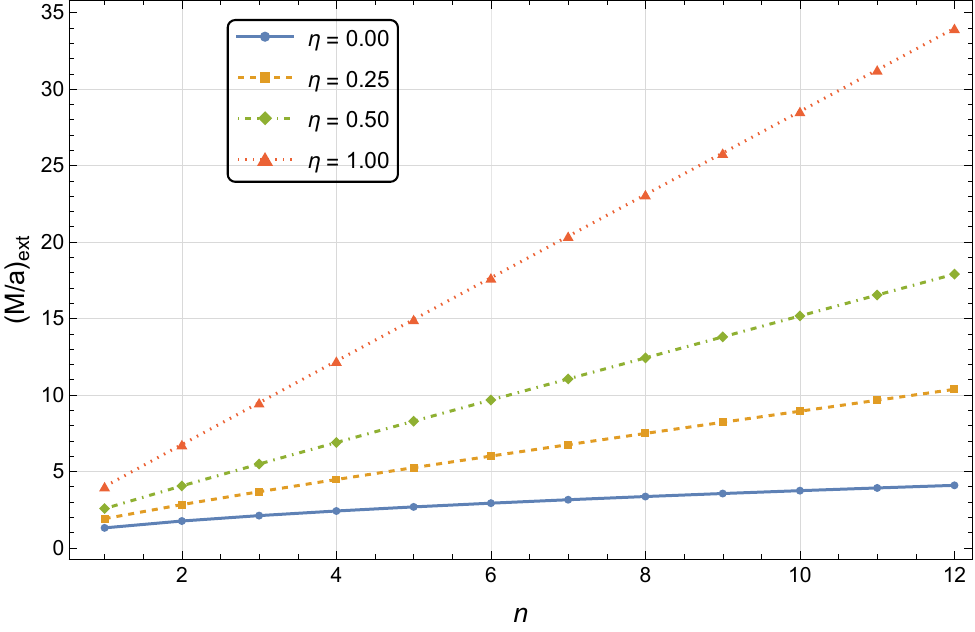}
 \caption{Extremal mass-to-length ratio $(M/a)_{\rm ext}$ versus the core index $n$ for several values of $\eta$. The increase of the critical ratio with $n$ shows explicitly why horizon existence imposes an upper admissible integer $n_{\max}$ once $M/a$ and $\eta$ are fixed.}
 \label{fig:extremality}
\end{figure}

The dependence of this constraint on the two deformation parameters is shown in Fig.~\ref{fig:extremality}. The critical ratio $(M/a)_{\rm ext}$ grows monotonically with $n$, and a nonzero $\eta$ steepens this growth substantially. Consequently, at fixed $M/a$ a horizontal line intersects each curve at a finite value of $n$, providing a direct graphical interpretation of the upper admissible core index.

\section{Horizon thermodynamics}

The thermodynamic quantities can now be expressed in terms of the outer horizon $r_+$. For a static metric of the form (1), the surface gravity is $\kappa=f'(r_+)/2$, and the Hawking relation $T_H=\kappa/(2\pi)$ \cite{BardeenCarterHawking1973,Hawking1975} gives
\begin{equation}
 T_H=\frac{f'(r_+)}{4\pi}.
\end{equation}
Differentiating the lapse (4) and then using the horizon relation (16) to eliminate $M$ produces the more useful horizon form
\begin{equation}
 T_H=\frac{s_+^2-2n\eta a s_+-(2n+1)a^2}
 {4\pi r_+s_+(s_++\eta a)}.
\end{equation}
The numerator in Eq.~(26) is precisely the dimensional version of the extremality polynomial (19). Therefore, at $r_+=r_e$, Eq.~(19) implies $T_H(r_e)=0$. In the opposite Schwarzschild limit $a\to0$, one has $r_+\to2M$ and Eq.~(26) reduces to $T_H\to1/(8\pi M)$.

\begin{figure}[!b]
 \centering
 \includegraphics[width=0.96\columnwidth]{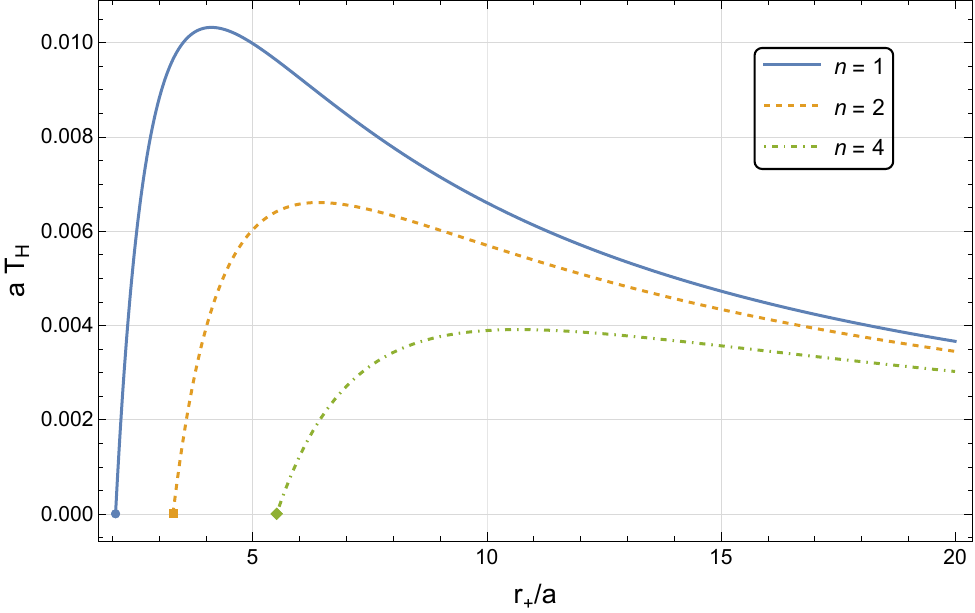}
 \caption{Dimensionless Hawking temperature $aT_H$ versus $r_+/a$ for $\eta=0.5$. The circles mark the corresponding extremal radii, where $T_H=0$. Larger $n$ moves the extremal configuration to larger radius and reduces the peak temperature.}
 \label{fig:temperature}
\end{figure}

Figure~\ref{fig:temperature} displays the dimensionless temperature $aT_H$ as a function of $r_+/a$ for $\eta=0.5$. Each branch starts at its own extremal radius with zero temperature, rises to a finite maximum, and then decreases for large horizons. Increasing $n$ shifts the extremal point outward and lowers the maximum temperature, making the thermodynamic effect of the core index directly visible.

Since the gravitational sector is ordinary Einstein gravity, the entropy is fixed by the Bekenstein--Hawking area law \cite{Bekenstein1973,Hawking1975}. With $A_+=4\pi r_+^2$,
\begin{equation}
 S=\frac{A_+}{4}=\pi r_+^2,
\end{equation}
and Eq.~(22) gives $S_e=\pi a^2(y_e^2-1)$ for the extremal state.

The horizon thermodynamic interpretation requires some care because $M$ is the ADM mass of an extended effective source, whereas the mass contained inside the horizon is the Misner--Sharp energy. The horizon equation (16), together with $f=1-2m/r$, gives directly
\begin{equation}
 E_+=m(r_+)=\frac{r_+}{2}.
\end{equation}
The radial Einstein equation evaluated at the horizon can be written as
\begin{equation}
 f'(r_+)=\frac{1}{r_+}+8\pi r_+P_+,
 \qquad P_+=p_r(r_+)=-\rho(r_+).
\end{equation}
Multiplying Eq.~(29) by $dr_+/(4\pi)$ in the combination appropriate to $T_HdS$, and using $dS=2\pi r_+dr_+$, $dE_+=dr_+/2$, and $dV_+=4\pi r_+^2dr_+$ with $V_+=4\pi r_+^3/3$, gives the horizon thermodynamic identity
\begin{equation}
 dE_+=T_H\,dS-P_+\,dV_+.
\end{equation}
Here $P_+=p_r(r_+)$ is the radial pressure of the effective matter source at the horizon; it is not the cosmological pressure $-\Lambda/(8\pi)$ used in extended black-hole thermodynamics. This rewriting of the horizon Einstein equation as a thermodynamic identity is closely related to the horizon-thermodynamics formulation developed in Refs.~\cite{Padmanabhan2002,Paranjape2006}. Thus Eq.~(30) is not postulated; it is the radial Einstein equation evaluated at the horizon and rewritten thermodynamically. Because the source extends outside $r_+$, in general $M\ne E_+$ and the exterior contribution is $M_{\rm out}=M-r_+/2$.

A complementary differential identity describes variations of the global ADM parameter. From the mass function in Eq.~(4), define at the horizon
\begin{equation}
 \sigma_+=\left[\frac{r_+}{s_++\eta a}\right]^{2n+1},
\end{equation}
so that the horizon equation is $M\sigma_+=r_+/2$. Taking the total differential of the mass relation (17), while keeping the discrete label $n$ fixed, and using $dS=2\pi r_+dr_+$ together with Eq.~(26), gives
\begin{equation}
 \sigma_+\,dM=T_H\,dS+\Phi_a\,da+\Phi_\eta\,d\eta,
\end{equation}
where the coefficients multiplying the independent variations are
\begin{equation}
 \Phi_a=\frac{(2n+1)r_+(a+\eta s_+)}{2s_+(s_++\eta a)},
 \qquad
 \Phi_\eta=\frac{(2n+1)ar_+}{2(s_++\eta a)}.
\end{equation}
Equations (30) and (32) have different interpretations: the former is the local horizon thermodynamic identity with $E_+=r_+/2$, while the latter is a parameter-space identity tracking variations of the global ADM mass in an extended-source geometry and should not be confused with an ordinary vacuum first law $dM=T_HdS$.

\section{Null geodesics and optical properties}

We next examine massless probes. Spherical symmetry allows the motion to be restricted to the equatorial plane, $\theta=\pi/2$. Because $t$ and $\phi$ are cyclic coordinates, the corresponding conserved quantities may be written directly as $E=f(r)\dot t$ and $L=r^2\dot\phi$. We denote the null effective potential by $V_{\rm eff}^{\rm N}$, where the uppercase superscript $\rm N$ labels the null sector and avoids confusion with the core index $n$. Inserting the conserved quantities into the null normalization condition $g_{\mu\nu}\dot x^\mu\dot x^\nu=0$ gives the standard one-dimensional null-geodesic form used in classic analyses of photon capture and black-hole optics \cite{Darwin1959,Synge1966}
\begin{equation}
 \dot r^2+V_{\rm eff}^{\rm N}(r)=E^2,
 \qquad
 V_{\rm eff}^{\rm N}(r)=\frac{L^2f(r)}{r^2}.
\end{equation}
This form makes the optical interpretation transparent: the radial motion is governed by the single potential $f(r)/r^2$. Defining the impact parameter $b=L/E$, a turning point $r_0$ satisfies $\dot r=0$ and therefore $b^2=r_0^2/f(r_0)$. Large enough $b$ corresponds to scattering, while the critical value is reached when the turning point becomes a circular unstable null orbit.

A circular photon orbit is therefore a degenerate turning point: in addition to $\dot r=0$, the radial potential must be stationary. Equivalently, $d[f(r)/r^2]/dr=0$, which gives the universal photon-sphere condition
\begin{equation}
 rf'(r)-2f(r)=0.
\end{equation}
The shadow-generating photon sphere is the outer solution of this equation that lies outside the event horizon and corresponds to a maximum of $V_{\rm eff}^{\rm N}$; equivalently, it satisfies $V_{\rm eff}^{\rm N\,\prime\prime}(r_{\rm ph})<0$.

Substitution of the lapse (4) into Eq.~(35) yields the exact implicit equation
\begin{equation}
\begin{aligned}
 Mr_{\rm ph}^{2n}\Big[&3s_{\rm ph}^2-2(n-1)\eta a s_{\rm ph}-(2n+1)a^2\Big]\\
 &=s_{\rm ph}(s_{\rm ph}+\eta a)^{2n+2},
\end{aligned}
\end{equation}
where $s_{\rm ph}=\sqrt{r_{\rm ph}^2+a^2}$. For generic $n$ this equation is not algebraically invertible and the physically relevant outer root is obtained numerically.

\begin{figure}[t]
 \centering
 \includegraphics[width=0.96\columnwidth]{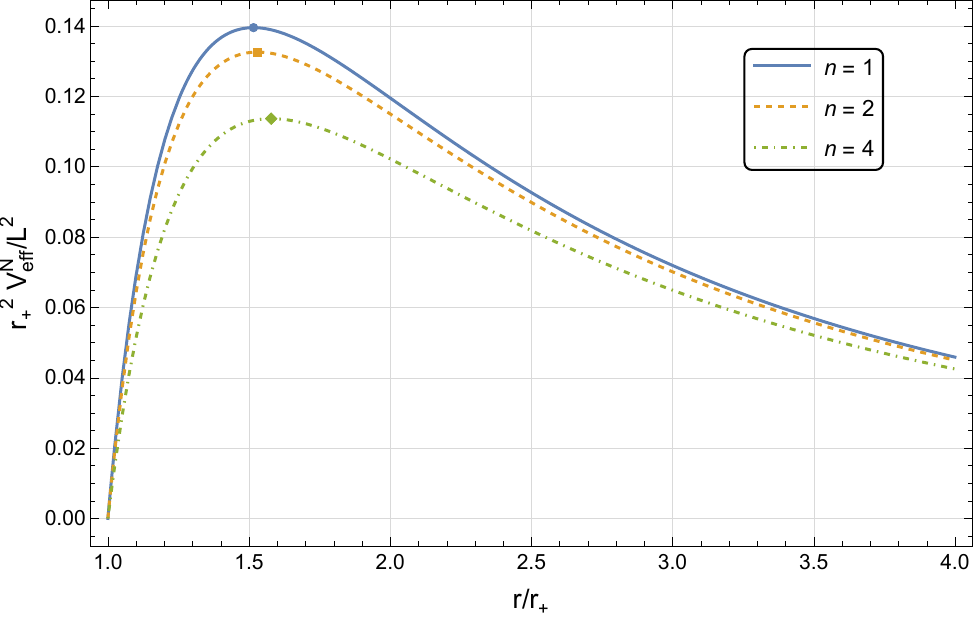}
 \caption{Normalized null effective potential $r_+^2V_{\rm eff}^{\rm N}/L^2$ as a function of $r/r_+$ for $M/a=10$, $\eta=0.5$, and representative values of $n$. Each curve begins at the event horizon $r/r_+=1$. The circles identify the outer unstable photon orbits determined by Eq.~(35).}
 \label{fig:nullpotential}
\end{figure}

The corresponding potential structure is shown in Fig.~\ref{fig:nullpotential} for the representative black-hole ratio $M/a=10$ and $\eta=0.5$. To compare solutions with different horizon radii on the same scale, we plot $r_+^2V_{\rm eff}^{\rm N}/L^2$ against $r/r_+$. Every curve begins at zero at the event horizon, reaches a single exterior maximum at the marked photon orbit, and then decays toward zero. For these representative parameters, increasing $n$ shifts the maximum slightly outward relative to the horizon and reduces its height.
 Once that root is known, Eq.~(36) can be used to eliminate $M$ from the lapse itself, giving
\begin{equation}
 f(r_{\rm ph})=
 \frac{s_{\rm ph}^2-2n\eta a s_{\rm ph}-(2n+1)a^2}
 {3s_{\rm ph}^2-2(n-1)\eta a s_{\rm ph}-(2n+1)a^2}.
\end{equation}
The critical impact parameter follows from the turning-point relation stated above evaluated at $r_0=r_{\rm ph}$, in direct analogy with the classical Schwarzschild capture problem \cite{Darwin1959,Synge1966}. Inserting Eq.~(37) gives
\begin{equation}
 b_c=r_{\rm ph}\sqrt{
 \frac{3s_{\rm ph}^2-2(n-1)\eta a s_{\rm ph}-(2n+1)a^2}
 {s_{\rm ph}^2-2n\eta a s_{\rm ph}-(2n+1)a^2}}.
\end{equation}
For an observer at infinity the shadow radius on the impact-parameter plane is therefore $R_{\rm sh}=b_c$, consistent with the standard optical construction of black-hole images \cite{Synge1966,Luminet1979}. For a static observer at a finite radius $r_o>r_+$, projection onto the local orthonormal frame gives instead
\begin{equation}
 \sin\vartheta_{\rm sh}=\frac{b_c\sqrt{f(r_o)}}{r_o}.
\end{equation}

To obtain analytic intuition when the regularization scale is small compared with the mass, take fixed finite $n$ and $\eta$, set $\alpha=a/M$ and solve Eq.~(36) perturbatively with the ansatz $r_{\rm ph}/M=3+c_1\alpha+c_2\alpha^2+\mathcal O(\alpha^3)$. Matching equal powers of $\alpha$ gives
\begin{equation}
\begin{aligned}
 \frac{r_{\rm ph}}{M}
 &=3-\frac{4(2n+1)\eta}{3}\alpha
 -\frac{5(2n+1)}{18}\alpha^2\\
 &\quad-\frac{(2n+1)(17n+1)}{27}\eta^2\alpha^2
 +\mathcal O(\alpha^3).
\end{aligned}
\end{equation}
Substituting this expansion into the exact impact parameter (38) and expanding to the same order gives
\begin{equation}
\begin{aligned}
 \frac{b_c}{3\sqrt3M}
 &=1-\frac{(2n+1)\eta}{3}\alpha\\
 &\quad-\frac{(2n+1)[(8n+1)\eta^2+3]}{54}\alpha^2
 +\mathcal O(\alpha^3).
\end{aligned}
\end{equation}
The limits follow immediately from Eqs.~(40)--(41). For the generalized Bardeen branch $\eta=0$,
\begin{equation}
\begin{aligned}
 \frac{r_{\rm ph}}{M}
 &=3-\frac{5(2n+1)}{18}\left(\frac{a}{M}\right)^2+\cdots,\\
 \frac{b_c}{3\sqrt3M}
 &=1-\frac{2n+1}{18}\left(\frac{a}{M}\right)^2+\cdots.
\end{aligned}
\end{equation}
The ordinary Bardeen case follows by setting $n=1$, while $a\to0$ gives the Schwarzschild values $r_{\rm ph}=3M$ and $b_c=3\sqrt3M$. For $\eta>0$, the linear terms agree with the small-charge RN expressions after the asymptotic identification $Q^2\to Q_{\rm eff}^2$ from Eq.~(14).

\section{Photon-orbit frequency and instability}

The photon-sphere radius and critical impact parameter characterize where the unstable null orbit is located and how it appears on the observer's impact-parameter plane. They do not, however, fully describe how rapidly a photon moves around that orbit or how quickly a nearby null trajectory departs from it. Two complementary coordinate-time diagnostics are therefore useful: the angular frequency $\Omega_{\rm ph}$ and the Lyapunov exponent $\lambda_{\rm ph}$. For circular null geodesics in static spherical spacetimes, these quantities provide a compact characterization of the orbital timescale and instability timescale \cite{Cardoso2009}. Using the conserved quantities introduced in Sec.~V and Eq.~(38), the photon-orbit frequency can be written as
\begin{equation}
 \Omega_{\rm ph}=\left.\frac{d\phi}{dt}\right|_{r_{\rm ph}}
 =\frac{\sqrt{f_{\rm ph}}}{r_{\rm ph}}=\frac{1}{b_c},
 \label{eq:omega-ph}
\end{equation}
where $f_{\rm ph}=f(r_{\rm ph})$. Thus $\Omega_{\rm ph}$ is already encoded in the shadow scale and provides a dynamical interpretation of the critical impact parameter.

The additional information comes from the instability of the circular orbit. A small radial perturbation behaves locally as $\delta r\propto e^{\pm\lambda_{\rm ph}t}$, and the coordinate-time Lyapunov exponent is \cite{Cardoso2009}
\begin{equation}
 \lambda_{\rm ph}=
 \sqrt{\frac{f_{\rm ph}}{2r_{\rm ph}^{2}}
 \left(2f_{\rm ph}-r_{\rm ph}^{2}f''_{\rm ph}\right)}.
 \label{eq:lambda-ph}
\end{equation}
Unlike $\Omega_{\rm ph}$, this quantity depends on the second radial derivative of the lapse and therefore probes the local curvature of the null effective potential at its maximum. Dividing the instability rate by the orbital frequency defines the dimensionless diagnostic
\begin{equation}
 \mathcal R_{\rm ph}\equiv\frac{\lambda_{\rm ph}}{\Omega_{\rm ph}}
 =\sqrt{\frac{2f_{\rm ph}-r_{\rm ph}^{2}f''_{\rm ph}}{2}}.
 \label{eq:Rph}
\end{equation}
The orbital period is $T_{\rm ph}=2\pi/\Omega_{\rm ph}$, so a perturbation accumulates a factor $\exp(2\pi\mathcal R_{\rm ph})$ over one coordinate orbital period. Consequently, $\mathcal R_{\rm ph}$ measures the instability of the light ring relative to its orbital motion rather than in absolute coordinate time. Schwarzschild gives $M\Omega_{\rm ph}=M\lambda_{\rm ph}=1/(3\sqrt{3})$ and hence $\mathcal R_{\rm ph}=1$.

The weak-deformation expansion makes the different sensitivities of the two diagnostics explicit. Using the same $\alpha=a/M$ introduced in Sec.~V and writing $p=2n+1$, substitution of the photon-radius expansion (40) into Eqs.~\eqref{eq:omega-ph}--\eqref{eq:lambda-ph} gives
\begin{equation}
\begin{aligned}
 M\Omega_{\rm ph}
 &=\frac{1}{3\sqrt3}\Bigg[1+\frac{p\eta}{3}\alpha
 +\frac{p}{18}\alpha^2\\
 &\hspace{1.2cm}+\frac{p(20n+7)}{54}\eta^2\alpha^2
 +\mathcal O(\alpha^3)\Bigg],\\
 M\lambda_{\rm ph}
 &=\frac{1}{3\sqrt3}\Bigg[1+\frac{p\eta}{9}\alpha
 -\frac{p}{27}\alpha^2\\
 &\hspace{1.2cm}+\frac{p(1-2n)}{54}\eta^2\alpha^2
 +\mathcal O(\alpha^3)\Bigg].
\end{aligned}
\label{eq:omega-lambda-exp}
\end{equation}
Their ratio is correspondingly
\begin{equation}
 \mathcal R_{\rm ph}
 =1-\frac{2p\eta}{9}\alpha
 -\frac{p\left[5+2(7n+1)\eta^2\right]}{54}\alpha^2
 +\mathcal O(\alpha^3).
 \label{eq:Rph-exp}
\end{equation}
These expressions show that the frequency and instability need not move in parallel. On the generalized Bardeen branch $\eta=0$, the leading correction is quadratic: increasing $n$ raises $\Omega_{\rm ph}$ but lowers $\lambda_{\rm ph}$ and therefore reduces $\mathcal R_{\rm ph}$. This comparison isolates the effect of the tunable core because the RN-like $1/r^2$ asymptotic term is absent. By contrast, for $n=1$ and $\eta>0$, Eq.~\eqref{eq:Rph-exp} already changes linearly, $\mathcal R_{\rm ph}=1-(2\eta/3)(a/M)+\cdots$, reflecting the new asymptotic sector generated by $\eta$.

\begin{figure}[t]
 \centering
 \includegraphics[width=0.96\columnwidth]{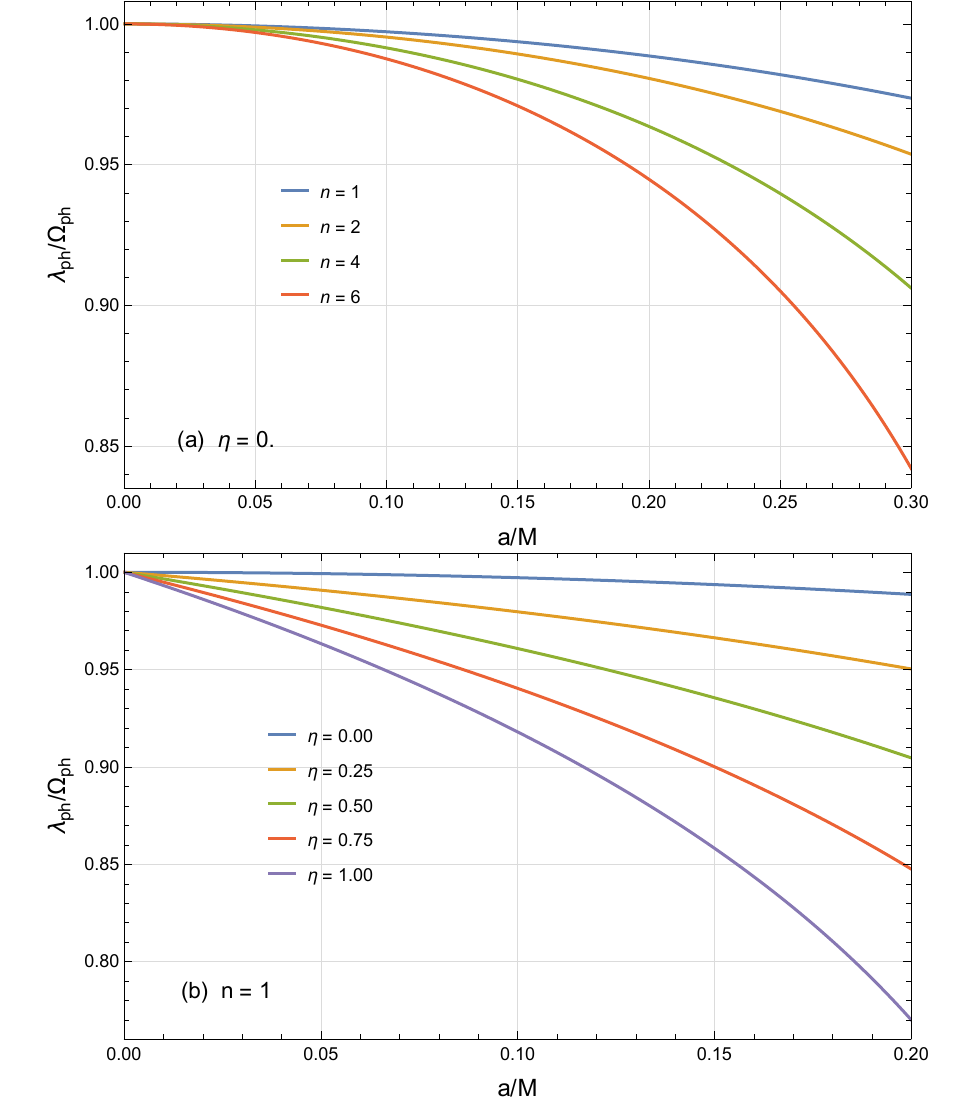}
 \caption{Exact photon-orbit instability ratio $\mathcal R_{\rm ph}=\lambda_{\rm ph}/\Omega_{\rm ph}$. (a) Generalized Bardeen branch $\eta=0$ for several core indices $n$, isolating the effect of the tunable central profile. (b) Lowest member $n=1$ for several values of $\eta$, showing the additional influence of the exterior deformation. The Schwarzschild limit corresponds to $\mathcal R_{\rm ph}=1$ at $a/M=0$. All displayed curves remain on the black-hole branch over the plotted intervals.}
 \label{fig:lyapratio}
\end{figure}

Figure~\ref{fig:lyapratio} shows the exact numerical ratio for these two complementary comparisons. Panel (a) fixes $\eta=0$ and varies $n$, thereby separating the core-index effect from the exterior RN-like deformation. At $a/M=0.30$, increasing $n$ from 1 to 6 changes $M\Omega_{\rm ph}$ from approximately $0.1955$ to $0.2086$, while $M\lambda_{\rm ph}$ decreases from approximately $0.1903$ to $0.1756$; the ratio consequently falls from $0.974$ to $0.842$. The light ring therefore revolves faster while becoming less unstable relative to its orbital period. This information cannot be inferred from the shadow radius alone and shows that different regular-core profiles can have distinguishable local photon-orbit dynamics even when their asymptotic behavior is of the same Schwarzschild type.

Panel (b) instead fixes $n=1$ and varies $\eta$, directly testing how the deformation of the ordinary Bardeen member changes the light-ring dynamics. At $a/M=0.20$, increasing $\eta$ from 0 to 1 raises $M\Omega_{\rm ph}$ from about $0.1938$ to $0.2544$, whereas $M\lambda_{\rm ph}$ changes only from about $0.1916$ to $0.1959$ and is mildly nonmonotonic, reaching a maximum at an intermediate $\eta$. The ratio nevertheless decreases strongly, from $0.989$ to $0.770$. Thus $\eta$ primarily shortens the orbital timescale, while the absolute instability rate is considerably less sensitive. In this sense the deformation changes not only the size of the shadow but also the balance between orbital motion and radial instability.

The distinction between $\Omega_{\rm ph}$ and $\lambda_{\rm ph}$ also has a direct perturbative interpretation. In the eikonal regime of asymptotically flat static spherical black holes, the real and imaginary parts of quasinormal frequencies are governed by the orbital frequency and Lyapunov exponent of the unstable null orbit \cite{Cardoso2009}. The unequal shifts found above therefore indicate that the core index $n$ and deformation $\eta$ can affect oscillation and damping scales differently. A dedicated quasinormal-mode calculation lies beyond the present geodesic analysis, but the photon-orbit diagnostics already identify which parameter combinations are expected to produce the clearest dynamical separation.

\section{Weak lensing and magnification}

The null radial equation above, together with $b=L/E$ and the turning-point relation $b^2=r_0^2/f(r_0)$, determines the photon trajectory. Integrating the azimuthal change for a ray arriving from infinity, reaching $r_0$, and returning to infinity and subtracting the flat-space value $\pi$ gives the exact deflection angle, in the standard form used in black-hole lensing analyses \cite{VirbhadraEllis2000,Bozza2002}
\begin{equation}
 \hat\alpha(b)=2\int_{r_0}^{\infty}
 \frac{b\,dr}{r^2\sqrt{1-b^2f(r)/r^2}}-\pi,
\label{eq:deflection-exact}
\end{equation}
with $b^2=r_0^2/f(r_0)$ as stated above. For generic $n$ and $\eta$, Eq.~\eqref{eq:deflection-exact} is evaluated numerically.

For weak lensing, the large-$r$ expansion (13) can be used to isolate the leading deformation sectors around the Schwarzschild lens. Systematic weak-field expansions for static spherical lenses and their observable corrections are developed, for example, in Refs.~\cite{KeetonPetters2005,KeetonPetters2006}. Here we do not attempt a complete post-Minkowskian expansion through order $b^{-3}$; instead, we retain the leading Schwarzschild term and separately display the pieces generated by the $1/r^2$ and $1/r^3$ deformations of our metric,
\begin{equation}
 \begin{aligned}
 \hat\alpha(b)&=\hat\alpha_{\rm Schw}(b)+\Delta\hat\alpha_{\rm def}(b),\\
 \Delta\hat\alpha_{\rm def}&\simeq-\frac{3\pi Q_{\rm eff}^2}{4b^2}-\frac{8C_3}{3b^3}+\cdots .
 \end{aligned}
\label{eq:deflection-def}
\end{equation}
where the $1/r^3$ coefficient of the lapse is
\begin{equation}
 C_3=M(2n+1)\left[1-(2n+2)\eta^2\right]a^2.
\label{eq:C3}
\end{equation}
The first term in $\Delta\hat\alpha_{\rm def}$ is the leading RN-like correction. The $C_3$ term is the contribution generated directly by the $1/r^3$ sector and reduces to the Bardeen-type correction when $\eta=0$. At the same order in $1/b$, a complete post-Minkowskian expansion also contains the usual Schwarzschild $M^3/b^3$ term and mixed pieces such as $M Q_{\rm eff}^2/b^3$; these are deliberately not included here. Accordingly, the analytic lens equation below is a sector-isolated leading-deformation model, useful for displaying how the $1/r^2$ and $1/r^3$ sectors enter the lens mapping, rather than a complete expansion to cubic order in $1/b$.

In the thin-lens approximation, $b\simeq D_L\theta$ and the angular lens equation takes the usual axisymmetric form \cite{VirbhadraEllis2000,KeetonPetters2005}
\begin{equation}
 \beta_s=\theta-\frac{D_{LS}}{D_S}\hat\alpha(D_L\theta).
\label{eq:lens-general}
\end{equation}
Defining the Einstein angle $\theta_E^2=4MD_{LS}/(D_LD_S)$ and combining the leading Schwarzschild baseline with the two isolated deformation terms in Eq.~\eqref{eq:deflection-def} gives
\begin{equation}
 \beta_s=\theta-\frac{\theta_E^2}{\theta}
 +\frac{A_2}{\theta^2}+\frac{A_3}{\theta^3},
\label{eq:lens-approx}
\end{equation}
with
\begin{equation}
 A_2=\frac{3\pi Q_{\rm eff}^2D_{LS}}{4D_SD_L^2},
 \qquad
 A_3=\frac{8C_3D_{LS}}{3D_SD_L^3}.
\label{eq:A2A3}
\end{equation}
For an axially symmetric lens, the signed Jacobian of the mapping between source and image angles gives the absolute magnification \cite{VirbhadraEllis2000,KeetonPetters2005}
\begin{equation}
 \mu=\left|\frac{\beta_s}{\theta}\frac{d\beta_s}{d\theta}\right|^{-1}.
\label{eq:mu-jac}
\end{equation}
Using Eq.~\eqref{eq:lens-approx} in Eq.~\eqref{eq:mu-jac} yields
\begin{equation}
\begin{aligned}
 \mu(\theta)\simeq\Bigg|&\left(1-\frac{\theta_E^2}{\theta^2}
 +\frac{A_2}{\theta^3}+\frac{A_3}{\theta^4}\right)\\
 &\times\left(1+\frac{\theta_E^2}{\theta^2}
 -\frac{2A_2}{\theta^3}-\frac{3A_3}{\theta^4}\right)\Bigg|^{-1}.
\end{aligned}
\label{eq:mu-theta}
\end{equation}

It is useful to display the observable limits more explicitly. With $x=\theta/\theta_E$, $y=\beta_s/\theta_E$, $\epsilon_2=A_2/\theta_E^3$, and $\epsilon_3=A_3/\theta_E^4$, and taking $y>0$ without loss of generality for the unresolved absolute magnification, Eq.~\eqref{eq:lens-approx} becomes
\begin{equation}
 y=x-\frac{1}{x}+\frac{\epsilon_2}{x^2}+\frac{\epsilon_3}{x^3}.
\label{eq:lens-dimless}
\end{equation}
Expanding the two image solutions and their Jacobians to first order in $\epsilon_2$ and $\epsilon_3$, the unresolved total magnification is
\begin{equation}
 \mu_{\rm tot}\simeq
 \frac{y^2+2}{y\sqrt{y^2+4}}
 +\frac{2\epsilon_2}{(y^2+4)^{3/2}}
 +\frac{4\epsilon_3}{y(y^2+4)^{3/2}}.
\label{eq:mu-total}
\end{equation}
Thus Schwarzschild follows from $\epsilon_2=\epsilon_3=0$,
\begin{equation}
 \mu_{\rm Schw}=\frac{y^2+2}{y\sqrt{y^2+4}}.
\label{eq:mu-schw}
\end{equation}
Within this sector-isolated truncation, Eq.~\eqref{eq:mu-total} displays separately the Schwarzschild baseline and the leading corrections generated by the $1/r^2$ and $1/r^3$ sectors. The $\epsilon_2$ term is RN-like because $\epsilon_2\propto Q_{\rm eff}^2$, whereas the $\epsilon_3$ term is core-sensitive and becomes the Bardeen-type contribution on the $\eta=0$ branch. For general $\eta$, $C_3$ also carries the mixed $\eta^2$ dependence. These terms have different source-position dependence, illustrating how the two deformation sectors enter the magnification. Because the omitted post-Minkowskian pieces contribute at the same cubic order in $1/b$, Eq.~\eqref{eq:mu-total} should not be interpreted as the complete magnification through that order.

\section{Timelike geodesics}

For a massive test particle, the equatorial geodesic Lagrangian follows from Eq.~(1) in the standard geodesic treatment of spherical black-hole spacetimes \cite{Darwin1959,PageThorne1974} as
\begin{equation}
 2\mathcal L=-f(r)\dot t^2+\frac{\dot r^2}{f(r)}+r^2\dot\phi^2=-1,
\label{eq:timelike-L}
\end{equation}
where the final equality is the timelike normalization for unit rest mass. Since $t$ and $\phi$ are cyclic coordinates, their Euler--Lagrange equations immediately give the conserved specific energy $E=f(r)\dot t$ and angular momentum $L=r^2\dot\phi$. We denote the timelike effective potential by $V_{\rm eff}^{t}$, where the superscript $t$ labels the timelike sector. Substituting these constants into the normalization condition in Eq.~\eqref{eq:timelike-L} gives the radial energy equation
\begin{equation}
 \dot r^2+V_{\rm eff}^{t}(r)=E^2,
 \qquad
 V_{\rm eff}^{t}(r)=f(r)\left(1+\frac{L^2}{r^2}\right).
\label{eq:timelike-radial}
\end{equation}
The potential vanishes at the horizon and approaches unity at infinity. For suitable $L$, it develops an unstable maximum and a stable minimum; the latter disappears at the innermost stable circular orbit. Since $V_{\rm eff}^{t}(\infty)=1$, circular orbits with $E<1$ are bound, while $E=1$ defines the marginally bound threshold.

For a circular orbit at $r=r_c$, both $\dot r=0$ and $dV_{\rm eff}^{t}/dr=0$ must hold. Solving these two equations simultaneously for $E^2$ and $L^2$ gives the familiar circular-orbit relations \cite{Darwin1959,PageThorne1974,Pugliese2011RN}
\begin{equation}
 E_c^2=\frac{2f^2}{2f-rf'},
 \qquad
 L_c^2=\frac{r^3f'}{2f-rf'}.
\label{eq:EcLc}
\end{equation}
Using $\Omega_c=\dot\phi/\dot t=fL/(Er^2)$ and substituting Eq.~\eqref{eq:EcLc} then gives
\begin{equation}
 \Omega_c^2=\frac{f'}{2r}.
\label{eq:Omega-c}
\end{equation}
The denominator $2f-rf'$ in Eq.~\eqref{eq:EcLc} vanishes at the photon-sphere condition (35), showing explicitly why ordinary timelike circular orbits lie outside the relevant null circular orbit.

The marginally bound circular orbit is obtained by imposing $E_c=1$ in the first relation of Eq.~\eqref{eq:EcLc}. After a simple rearrangement one obtains the exact condition
\begin{equation}
 rf'=2f(1-f).
\label{eq:mb-cond}
\end{equation}
For the full lapse, Eq.~\eqref{eq:mb-cond} is solved numerically. To obtain an analytic weak-deformation result for $a/M\ll1$, expand the exact lapse consistently about the Schwarzschild orbit (equivalently use the small-$a/r$ form (13) near $r\simeq4M$) in Eq.~\eqref{eq:mb-cond} and use the ansatz $r_{\rm mb}/M=4+d_1(a/M)+d_2(a/M)^2+\cdots$. Matching successive powers of $a/M$ yields
\begin{equation}
\begin{aligned}
 \frac{r_{\rm mb}}{M}
 &=4-2(2n+1)\eta\frac{a}{M}\\
 &\quad-\frac{3(2n+1)(1+2n\eta^2)}{8}
 \left(\frac{a}{M}\right)^2+\cdots.
\end{aligned}
\label{eq:rmb-exp}
\end{equation}

Stability is determined by the curvature of the effective potential. At the limiting stable circular orbit, the conditions $V_{\rm eff}^{t\,\prime}=0$ and $V_{\rm eff}^{t\,\prime\prime}=0$ hold simultaneously. Eliminating $E$ and $L$ with Eq.~\eqref{eq:EcLc}, or equivalently differentiating $L_c^2(r)$ and setting $dL_c^2/dr=0$, gives the exact ISCO equation
\begin{equation}
 rff''+3ff'-2r(f')^2=0.
\label{eq:isco-cond}
\end{equation}
For the general lapse this equation is again solved numerically. For $a/M\ll1$, expanding the exact equation about the Schwarzschild value with $r_{\rm ISCO}/M=6+e_1(a/M)+e_2(a/M)^2+\cdots$ gives
\begin{equation}
\begin{aligned}
 \frac{r_{\rm ISCO}}{M}
 &=6-3(2n+1)\eta\frac{a}{M}\\
 &\quad-\frac{19(2n+1)(1+2n\eta^2)}{36}
 \left(\frac{a}{M}\right)^2+\cdots.
\end{aligned}
\label{eq:risco-exp}
\end{equation}

The Schwarzschild values follow directly by setting $a\to0$ in Eqs.~\eqref{eq:rmb-exp} and \eqref{eq:risco-exp}, giving $r_{\rm mb}^{\rm Sch}=4M$ and $r_{\rm ISCO}^{\rm Sch}=6M$, in agreement with the classic Schwarzschild orbit analysis \cite{Darwin1959}. The same two equations already display the physical origin of the corrections without requiring a second, nearly identical pair of formulas. The terms linear in $\eta a/M$ originate from the RN-like $1/r^2$ sector of the asymptotic lapse. In particular, using $Q_{\rm eff}^2=2M(2n+1)\eta a$, the leading ISCO shift becomes $-(3/2)Q_{\rm eff}^2/M^2$, consistent with the small-charge Reissner--Nordstr\"om result \cite{Pugliese2011RN}. The parts proportional to $(a/M)^2$ that remain when $\eta=0$ are the generalized-Bardeen core corrections; for $n=1$ they reduce to the ordinary Bardeen shifts. The additional pieces proportional to $\eta^2(a/M)^2$ describe the first nonlinear mixing between the exterior deformation and the regular-core scale and should not be identified with either a pure RN or a pure Bardeen contribution.

\begin{figure}[!b]
 \centering
 \includegraphics[width=0.96\columnwidth]{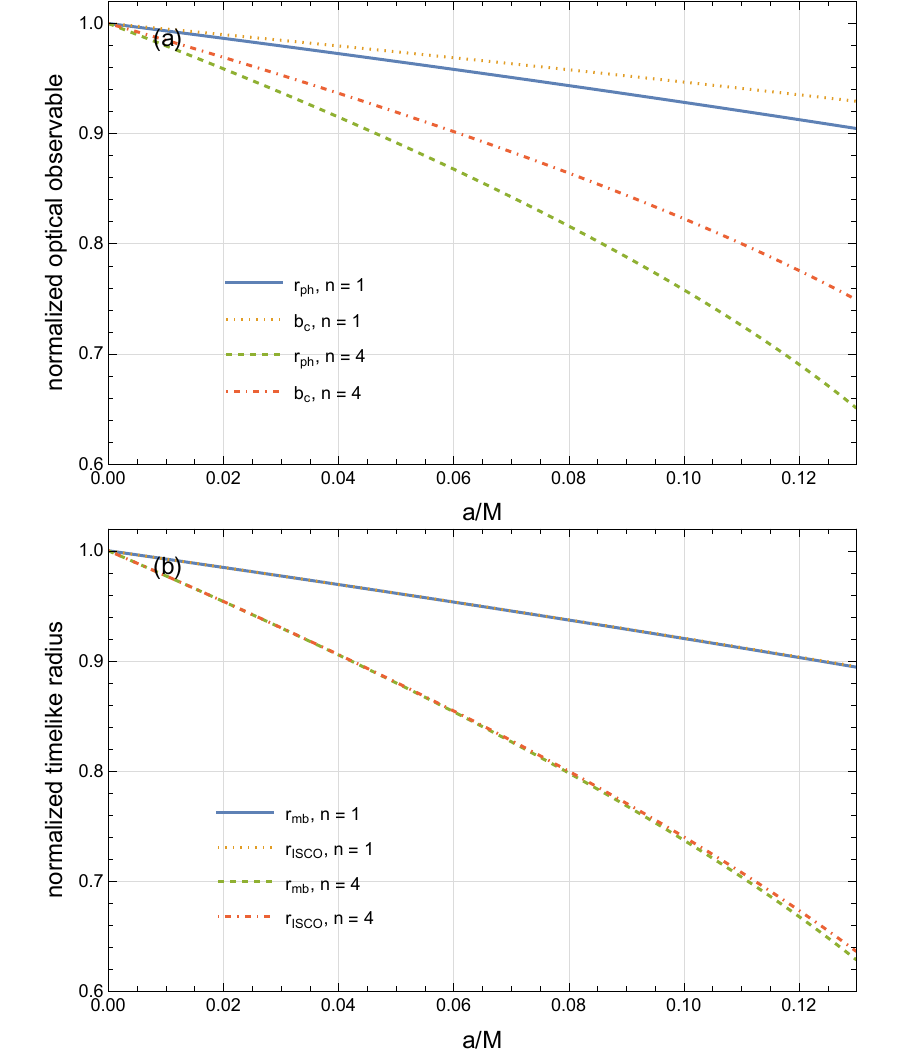}
 \caption{Exact numerical geodesic observables versus $a/M$ for $\eta=0.5$. (a) Photon radius $r_{\rm ph}/(3M)$ and critical impact parameter $b_c/(3\sqrt{3}M)$. (b) Marginally bound radius $r_{\rm mb}/(4M)$ and ISCO radius $r_{\rm ISCO}/(6M)$. The Schwarzschild limit corresponds to unity at $a/M=0$. The plotted interval remains on the black-hole branch for both $n=1$ and $n=4$.}
 \label{fig:geodesicobs}
\end{figure}

A direct comparison of the exact numerical geodesic observables is given in Fig.~\ref{fig:geodesicobs}. We fix $\eta=0.5$ and show the $n=1$ and $n=4$ branches over $0\le a/M\le0.13$, a range that remains inside the black-hole domain for both choices. Panel (a) normalizes the photon radius and critical impact parameter to their Schwarzschild values, while panel (b) does the same for the marginally bound and ISCO radii. All four observables approach unity as $a/M\to0$, providing a numerical check of the Schwarzschild limit, while the stronger downward shift for $n=4$ demonstrates how the tunable core affects both null and timelike geodesic scales.

For parameter values continuously connected to the Schwarzschild weak-deformation regime, the photon, marginally bound, and marginally stable radii can be compared after solving their respective defining equations. We therefore do not impose a universal ordering as an independent relation for arbitrary $n$, $\eta$, and $a/M$.

\section{Conclusions}

The main result of this work is not only the construction of another regular metric, but the separation of two physically different deformations within the same family. The discrete index $n$ controls the structure of the central source, whereas $\eta$ controls an independent exterior deformation. This distinction is already visible at the level of the density. The $n=1$ member has a finite central density and a de Sitter-type core, while every $n>1$ member has $\rho(0)=0$ and develops a density maximum away from the center, as shown in Fig.~\ref{fig:density}. Increasing $n$ therefore transfers the effective support outward and produces progressively more depleted, Minkowski-like cores. By contrast, switching on $\eta$ does not destroy regularity but changes the large-radius tail from $r^{-5}$ to $r^{-4}$ and generates the asymptotic scale $Q_{\rm eff}^2=2M(2n+1)\eta a$. In the particularly important $n=1$ sector, this continuously deforms the ordinary Bardeen geometry while lowering its central density and replacing the asymptotic tangential-DEC violation of the $\eta=0$ branch by the limiting Maxwell-like behavior $q\to4$. Thus $n$ and $\eta$ act on complementary regions of the geometry rather than representing redundant parameters.

A second general result is that regularity alone does not guarantee a black hole for arbitrarily large $n$. The horizon condition converts the core index into a dynamically restricted quantity. Figure~\ref{fig:extremality} shows that the extremal ratio $(M/a)_{\rm ext}$ grows monotonically with $n$, and that nonzero $\eta$ makes this growth substantially steeper. Consequently, once $M/a$ and $\eta$ are fixed, only a finite set of integers $n\le n_{\max}$ can support horizons. In this sense, stronger central depletion requires a correspondingly larger mass-to-core scale in order to maintain a black-hole configuration. The asymptotic estimates, $(M/a)_{\rm ext}\sim\sqrt{en/2}$ for $\eta=0$ and $(M/a)_{\rm ext}\sim e\eta n$ for fixed $\eta>0$, make this restriction particularly transparent.

The same competition leaves a clear thermodynamic imprint. The extremal radius is simultaneously the zero-temperature endpoint, and Fig.~\ref{fig:temperature} shows that increasing $n$ moves this endpoint to larger $r_+/a$ while lowering the maximum Hawking temperature. Hence more strongly depleted cores are not thermodynamically equivalent to the Bardeen-like member: at fixed $\eta$ they require a larger extremal configuration and exhibit a reduced temperature scale. The entropy remains the standard area entropy because the gravitational action is unchanged. The relation $dE_+=T_HdS-P_+dV_+$ should be understood as a horizon thermodynamic identity following from the radial Einstein equation, with $P_+=p_r(r_+)$ the radial matter pressure rather than a cosmological pressure. This distinction is important because the ADM mass includes the portion of the effective source lying outside the horizon, whereas $E_+=r_+/2$ is the quasi-local horizon energy.

The optical sector exhibits another useful separation between core and exterior effects. For the representative black-hole domain displayed in Fig.~\ref{fig:nullpotential}, the null potential possesses a single exterior unstable maximum, and changing $n$ shifts both the position and height of this barrier. The perturbative formulas sharpen this result: on the generalized Bardeen branch $\eta=0$, the first changes in $r_{\rm ph}$ and $b_c$ are quadratic in $a/M$, whereas $\eta\neq0$ generates corrections already linear in $a/M$. This provides a simple diagnostic of the origin of a deviation from Schwarzschild behavior: the pure core deformation enters more weakly, while the RN-like exterior sector appears one order earlier. The exact numerical comparison in Fig.~\ref{fig:geodesicobs} shows the same tendency in both null and timelike observables: as $a/M$ increases, the photon radius, shadow scale, marginally bound radius, and ISCO move below their Schwarzschild-normalized values, with the displacement becoming appreciably stronger for larger $n$.

The photon-orbit instability analysis provides the clearest new dynamical behavior of the model. The shadow determines $\Omega_{\rm ph}=1/b_c$, but it does not determine how rapidly nearby rays leave the unstable light ring. The latter information is carried by $\lambda_{\rm ph}$ and, more cleanly, by the ratio $\mathcal R_{\rm ph}=\lambda_{\rm ph}/\Omega_{\rm ph}$. Figure~\ref{fig:lyapratio} shows that these quantities do not vary in parallel. On the $\eta=0$ branch, increasing $n$ raises the photon-orbit frequency while reducing the Lyapunov exponent relative to the orbital rate. For example, at $a/M=0.30$ the exact ratio decreases from approximately $0.974$ for $n=1$ to $0.842$ for $n=6$. The light ring therefore rotates faster but is less unstable per orbital cycle as the core becomes more depleted. This behavior cannot be inferred from the shadow size alone. The $n=1$ comparison reveals a different effect: at $a/M=0.20$, increasing $\eta$ from $0$ to $1$ lowers $\mathcal R_{\rm ph}$ from about $0.989$ to $0.770$, while $\Omega_{\rm ph}$ increases strongly and $\lambda_{\rm ph}$ changes much more mildly and even becomes weakly nonmonotonic. Thus the Bardeen deformation primarily changes the orbital timescale, whereas the absolute instability rate is considerably less sensitive. In the eikonal correspondence, this unequal response suggests that the real and imaginary parts of high-frequency quasinormal modes need not be shifted in the same way by $n$ and $\eta$.

The timelike sector leads to the same qualitative hierarchy. The marginally bound orbit and the ISCO recover $4M$ and $6M$ in the Schwarzschild limit, while the leading $\eta$-dependent shifts originate from the RN-like $1/r^2$ sector and the terms surviving at $\eta=0$ encode the generalized-Bardeen core correction. The nonlinear $\eta^2a^2$ terms then measure the first mixing between these two sectors. The weak-lensing analysis exhibits an analogous decomposition into a Schwarzschild baseline, a leading RN-like contribution, and a core-sensitive contribution. Because the lensing expression retained here is deliberately sector-isolated rather than a complete cubic post-Minkowskian expansion, its main role is to expose this separation of physical effects rather than to provide the final high-precision lensing observable.

Taken together, the calculations and figures lead to a simple physical picture. The parameter $n$ controls \emph{how the regular core is emptied}, while $\eta$ controls \emph{how the exterior departs from the Bardeen/Schwarzschild sector}. Their effects can therefore be distinguished through horizon existence, the temperature scale, the order at which optical and orbital corrections appear, and especially the relative photon-orbit instability $\lambda_{\rm ph}/\Omega_{\rm ph}$. The model consequently interpolates between a finite-density Bardeen core and increasingly depleted regular cores without losing asymptotic flatness, while simultaneously allowing an independently tunable RN-like exterior. The remaining limitation is that the effective stress tensor has been treated phenomenologically rather than derived from a fundamental matter Lagrangian. A natural continuation is therefore to investigate whether the source admits a microscopic realization and to compute the full quasinormal-mode spectrum, for which the photon-orbit results obtained here provide concrete analytic expectations.

\end{document}